\documentclass[twocolumn,pre]{revtex4-1}
\usepackage[latin9]{inputenc}
\usepackage{amssymb}
\usepackage{graphicx}

\makeatletter

\providecommand{\tabularnewline}{\\}

\@ifundefined{textcolor}{}
{%
 \definecolor{BLACK}{gray}{0}
 \definecolor{WHITE}{gray}{1}
 \definecolor{RED}{rgb}{1,0,0}
 \definecolor{GREEN}{rgb}{0,1,0}
 \definecolor{BLUE}{rgb}{0,0,1}
 \definecolor{CYAN}{cmyk}{1,0,0,0}
 \definecolor{MAGENTA}{cmyk}{0,1,0,0}
 \definecolor{YELLOW}{cmyk}{0,0,1,0}
}

\makeatother

\begin{document}

\title{Jamming in Hierarchical Networks}

\author{Xiang Cheng and Stefan Boettcher}

\affiliation{Department of Physics, Emory University, Atlanta, GA 30322, USA}
\begin{abstract}
We study the Biroli-Mezard model for lattice glasses on a number of
hierarchical networks. These networks combine certain lattice-like features
with a recursive structure that makes them suitable for exact renormalization 
group studies and provide an alternative to the mean-field approach. In our 
numerical simulations here, we first explore their equilibrium 
properties with the Wang-Landau algorithm. Then, we investigate
their dynamical behavior using a grand-canonical annealing algorithm.
We find that the dynamics readily falls out of equilibrium and jams
in many of our networks with certain constraints on the neighborhood
occupation imposed by the Biroli-Mezard model, even in cases where
exact results indicate that no ideal glass transition exists. But
while we find that time-scales for the jams diverge, our simulations
can not ascertain such a divergence for a packing fraction  distinctly 
above random close packing. In cases where we allow hopping in our 
dynamical simulations, the jams on these networks generally disappear.
\end{abstract}
\maketitle

\section{Introduction}

\label{sec:intro} The jamming transition, as discussed by Liu and
Nagel in 1998~\cite{Liu1998}, for example, has been the focus of
intense study~\cite{Biroli2007,Liu2010}. A granular disordered system
for increasing density can reach a jammed state at which a finite
yield stress develops, or at least extremely long relaxation times
ensue, similar to the emerging sluggish behavior observed when the
viscosity of a cooled glassy liquid seemingly diverges. Thus, a jamming
transition may be induced in various ways, such as by increasing density,
decreasing temperature, or/and reducing shear stress~\cite{Liu2010}.
Below the jamming transition, the system stays in long-lived meta-stable
states, and its progression to its corresponding equilibrium state
entails an extremely slow, non-Debye relaxation~\cite{Hill1985,Ciamarra2010,van2010}.
Jamming transitions have been observed in various types of systems,
such as granular media~\cite{Majmudar2007}, molecular glasses~\cite{Parisi2010,Angelani2007},
colloids~\cite{Trappe2001}, emulsions~\cite{Zhang2005}, foams~\cite{Berthier2011,DaCruz2002},
etc~\cite{Liu2010,van2010}. These systems can behave like stiff solids
at a high density with low temperature and small perturbations. In
these transitional processes, the systems can self-organize their
own structure to avoid large fluctuations~\cite{Berthier2011} and
to reach a quasi-stable jammed state, characterized by an extremely
slow evolution to the unjammed equilibrium state. The properties of
those quasi-stable non-equilibrium states as well as their corresponding
equilibrium state is the main focus of this paper. 

The properties of the jamming transition have been studied 
extensively~\cite{Biroli2007,Majmudar2007,Liu2010}, but we still lack an essential
understanding of the physics underlying the jammed state. Theoretical
progress has been much slower than the accumulation of experimental
discoveries. One of the reasons is the scarcity of theoretical microscopic
models to capture the complex jamming process~\cite{Krzakala2008,Jacquin2011}.
In recent years, a lattice glass model proposed by Biroli and Mezard
(BM)~\cite{Biroli02} has been shown as a simple but adequate means
to study the jamming process. It is simple because the model follows
specific dynamical rules which are elementary to implement in both
simulations and analytical work. In distinction to kinetically constrained
models such as that due to Kob and Andersen~\cite{Kob93}, in which
particles are blocked from leaving a position unless certain neighborhood
conditions are satisfied, BM embeds geometric frustration merely by
preventing the neighborhood of any particles to consist of more than
$l$ other particles. Beyond that, it proceeds purely thermodynamically.
The phase diagram can be reduced to just one (or both) of two control
parameters, chemical potential and temperature. Either is sufficient
to reproduce a jamming transition which is similar to that observed
in off-lattice systems~\cite{Biroli02}. Using this model in a mean-field
network (i.e., a regular random graph), Krzakala \textit{et al.} find
jammed states in Monte Carlo simulations and a genuine thermodynamical
phase transition (ideal glass transition) in its mean-field analytical
solutions~\cite{Krzakala2008}. In other words, the jammed state coincides
with an underlying equilibrium state that possesses a phase transition
to a glassy state. That raises the prospect that this glass transition
might be the reason for the onset of jamming. The evidence for such
a connection thus far is based on mean-field models~\cite{Biroli02,Krzakala2008, Rivoire03},
as such a transition is hard to ascertain for finite-dimensional lattice
glasses. Yet, it remains unclear whether mean-field solutions in disordered
systems can provide an adequate conception for real-world behavior. 

In this paper, we propose to use the lattice glass model BM on hierarchical
networks~\cite{Boettcher2008HN}, which are networks with a fixed,
lattice-like geometry. They combine a finite-dimensional lattice backbone
with a hierarchy of small-world links that in themselves impose a
high degree of geometric frustration despite of their regular pattern.
In fact, the recursive nature of the pattern can ultimately provide
analytical solution via the renormalization group (RG), positioning
these networks as sufficiently simple to solve as well as sufficiently
lattice-like to become an alternative to mean-field solutions~\cite{BoHa11}. Unlike mean-field models, our network is dominated by many small loops that are also the hallmark of lattice systems.
Our goal is to find (1) whether the lattice glass model leads to jamming
state in hierarchical networks, (2) whether there is an ideal glass
transition underlying the jamming transition, and (3) whether the
local dynamics affect the jamming process. To our knowledge, these
questions have not been studied in any small-world systems. Our results
can contribute new insights to understand jamming.

We find that BM in these networks can jam, even when there is certifiably
no equilibrium transition; the geometric frustration that derives
from the incommensurability among the small-world links is sufficient
in many cases to affect jamming. In fact, jamming is most pronounced
for fully exclusive neighborhoods ($l=0$). It disappears for more
disordered neighborhoods ($l=1$), at least for our non-regular networks,
where the allowance of $l=1$ neighbor to be occupied seems to provide
the ``lubrication'' that averts jams. However, the packing fractions
at which time-scales diverge is virtually indistinguishable from random
close packing within the accuracy of our simulations. 

Mean-field calculations of BM in Ref.~\cite{Rivoire03} predict a kinetic transition for dynamic rules based on nearest-neighbor hopping. In our simulations, we find that  such hopping,  in addition to the particle exchange with a bath, can affect a dramatic change in the dynamic behavior and eliminated jamming in all cases we consider.  

This paper is organized as follows. In Sec.~\ref{sec:mm}, we describe
the model, the networks, and out numerical simulations. In Sec.~\ref{sec:results}
we discuss the results of our simulations for each network. In Sec.
\ref{sec:conclusions} we conclude with a few summary remarks and
an outlook for future work.

\section{Model \& Methods}

\label{sec:mm} In this section, we describe the model and the networks
on which we will study its behavior. To benchmark the equilibrium
properties of the model on those networks, we implement a multi-canonical
algorithm due to Wang and Landau~\cite{Wang2001,wang:01a}. We further
need a grand-canonical annealing algorithm to study the dynamics of the
lattice glass model on those networks.

\subsection{Lattice glass model}

\label{subsec:latticemodel} The lattice glass model as defined by
Biroli and Mezard (BM)~\cite{Biroli02} considers a system of particles
on a lattice of $N$ sites. Each site can carry either $x_{i}=0$
or $x_{i}=1$ particle, and the occupation is restricted by a hard,
local ``density constraint'': any occupied site ($x_{i}=1$) can
have at most $l$ occupied neighbors, where $l$ could range locally
from 0 to the total number of its neighbor-sites. In this model, the
jamming is defined thermodynamically by rejecting the configurations
violating the density constraint. Here, we focus on global density
constraints of $l=0$ (completely excluded neighborhood occupation)
and $l=1$ as the most generic cases. The system can be described
by the grand canonical partition function 
\begin{equation}
Z(\mu)=\sum_{{\bf allowed}\,\{x_{i}\}}\exp\left[\mu\sum_{i=1}^{N}x_{i}\right],\label{eq:PF}
\end{equation}
where the sum is over all the allowed configurations $\{x_{i}\}$.
Here, $\mu$ is the reduced chemical potential, where we have chosen
units such that the temperature is $k_{B}T=1/\beta=1$, and $\sum_{i=1}^{N}x_{i}$
is the total number of particles in a specific configuration.

From the grand canonical partition function in Eq.~(\ref{eq:PF}),
we can obtain the thermodynamic observables we intend to measure,
such as the Landau free energy density $w(\mu)$, the packing fraction
$\rho(\mu)$, and the entropy density $s\left[\rho(\mu)\right]$,
as defined in the following equations:
\begin{eqnarray}
w(\mu) & = & -\frac{1}{N}\ln Z,\label{eq:GCPdiff}\\
\rho(\mu) & = & \frac{1}{N}\left<\sum_{i=1}^{N}x_{i}\right>_{\mu}=\frac{1}{N}\frac{\partial\ln Z}{\partial\mu},\nonumber \\
s(\mu) & = & \frac{1}{N}\left(1-\mu\frac{\partial}{\partial\mu}\right)\ln Z.\nonumber 
\end{eqnarray}

\subsection{Hierarchical networks}

\label{subsec:HN} In our investigations, we use the Hanoi networks
~\cite{Boettcher2008HN}. These are small-world networks with a hierarchical,
recursive structure that avoid the usual randomness involved in defining
an ensemble of networks. Thus, no additional averages of such an ensemble
are required to obtain scaling properties in thermodynamical limit
from a finite system size, which reduces the computational effort.
Hanoi networks combine a real-world geometry with a hierarchy of small-world
links, as an instructive intermediary between mean-field and finite-dimensional
lattice systems, on which potentially exact results can be found using
the renormalization group~\cite{BoHa11}. 

\begin{figure}
\centering \includegraphics[width=1\columnwidth]{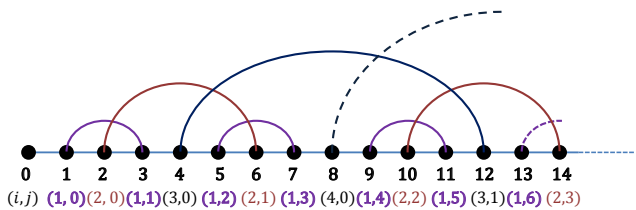} 
\protect\caption{\label{fig:HN3_short} An example of the first 14 sites of HN3 on a semi-infinite line.}
\end{figure}

We use three Hanoi networks: HN3 is a network with a regular degree
of 3, while HN5 is a similar network that possesses many extra links
such that vertices have an exponential arrangement of degrees with
an average degree of 5. HNNP is a similar Hanoi network of average
degree of 4, but which is non-planar. Each of them can be built on
a simple backbone of a 1D lattice. The 1D backbone has $N=2^{k}+1$
($k=1,2,3,\cdots$) sites where each site is numbered from $0$ to
$N$. Any site $n$, $0\le n\le N$, can be defined by two unique
integers $i$ and $j$, 
\begin{equation}
n(i,j)=2^{i-1}(2j+1),\label{eq:numbering}
\end{equation}
where $i$, $1\le i\le k$, denotes the level in the hierarchy and
$j$, $0\le j<2^{k-i}$, labels consecutive sites within each
hierarchy $i$. Site $n=0$ is defined in the highest level $k$ or,
equivalently, is identified with site $n=N$ for periodic boundary
conditions. With this setup, we have a 1D backbone of degree
2 for each site and a well-defined hierarchy on which we can build
long-range links recursively in three different ways: HN3~\cite{Boettcher2008HN}
is constructed by connecting the neighbor sites $n(i,0)\longleftrightarrow n(i,1)$,
$n(i,2)\longleftrightarrow n(i,3)$, $n(i,4)\longleftrightarrow n(i,5)$,
and so on and so forth. For example, in level $i=1$, site $n(1,0)=1$
is connected to $n(1,1)=3$; site $n(1,2)=5$ is connected to $n(1,3)=7$;
and so on. A initial section of a HN3 network is given in Fig.~\ref{fig:HN3_short}.
As a result, HN3 is a planar network of regular degree 3.

\begin{figure}
\centering\includegraphics[width=0.8\columnwidth]{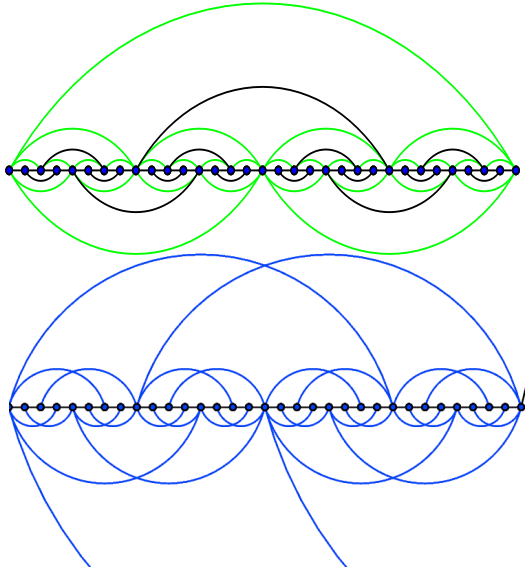}
\caption{\label{fig:Depiction-of-HN}Depiction of HN5 (top) and HNNP (bottom),
first introduced in Ref.~~\cite{Boettcher09c}. Green-shaded lines
in HN5 represent its difference to HN3, which is at its core (dark
lines). While HN3 and HN5 are planar, HNNP is non-planar. }
\end{figure}

HN5~\cite{Boettcher09c}, as shown in Fig.~\ref{fig:Depiction-of-HN},
is an extension based on HN3, where each site in level $i$ ($i\ge2$,
i.e., all even sites) is further connected to sites that are $2^{i-1}$
sites away in both directions. For example, for the level $i=2$ sites
(sites $2,6,10,\cdots$), site 2 is connected to both site 0 and site
4; site 6 is connected to sites 4 and 8; etc. The resulting network
remains planar but has a hierarchy-dependent degree, i.e., 1/2 sites
have degree 3, 1/4 have degree 5, 1/8 have degree 7, etc. In the limit
of $N\to\infty$, this network has  average degree 5.

HNNP~\cite{Boettcher09c}, also shown in Fig.~\ref{fig:Depiction-of-HN},
is constructed from the same 1D backbone as HN3 and HN5. However,
for site $n$ in level $i$ with even $j$, it is connected forwards
to site $(n+3\times2^{i-1})$; while site $n$ in level $i$ with
odd $j$ is connected backwards to site $(n-3\times2^{i-1})$. Level
1 and level 2 sites have degree 3, and level $3,4,5,\cdots$ sites
have degree $5,7,9,\cdots$. The HNNP has an average degree of 4 and
is non-planar.

\subsection{Wang-Landau Sampling}

\label{sub:WLsampling}Wang-Landau sampling~\cite{Wang2001} is a
multi-canonical method to numerically determine the entire density
of states $g_{n}$ within a single simulation. This method is based
on the fact that a random walk in the configuration space with a probability
proportional to the inverse of the density of states with occupation
$n$, $0\leq n\leq N$, enforces a flat histogram in $g_{n}$ over
all $n$. Based on this fact, Wang-Landau sampling keeps modifying
the estimated density of states in the random walks over all possible
configurations and can make the density of states converge to the
true value. The update procedure is: 
\begin{enumerate}
\item Initially, set all unknown density of states $\{g_{n}=1\}$ and the
histogram $\{H_{n}=0\}$ for all occupations $n$, initiate the modification
factor $f=e^1\approx2.71828\dots$ ; 
\item Randomly pick a site $i$; if it is empty (occupied), add (remove)
a particle with a probability of $\min\left[1,\frac{g_{n}}{g_{n+1}}\right]$
($\min\left[1,\frac{g_{n}}{g_{n-1}}\right]$) while obeying the rule
of the hard local density constraint on having at most $l$ occupied
nearest neighbors of site $i$; 
\item Randomly pick one occupied site and one empty site; transfer
a particle from the occupied site to the empty, if the density constraint is not violated; 
\item Update the $H_{n}$ and $g_{n}$ of the current state, i.e., set $\{H_{n}=H_{n}+1\}$
and $\{g_{n}=g_{n}\times f\}$; 
\item Repeat steps 2 to 4 until the sampling reaches a nearly flat histogram
for the $H_{n}$, then update the modification factor $f=\sqrt{f}$
and reset $\{H_{n}=0\}$; 
\item Stop if $f\le1+10^{-8}$. 
\end{enumerate}
Our procedure mostly follows the standard procedure of Wang-Landau
sampling~\cite{Wang2001}, except for step 3. Its purpose is to facilitate
the random walk to explore phase space more broadly and to expedite
convergence.

Wang-Landau sampling has been proved as an effective method to find
the density of states~\cite{Wang2001,Lee2006,Dickman2011}. In our
study, it can find convergence for system size of up to $N\sim10^{3}$
within a reasonably computational cost. From the density of states,
we can calculate the equilibrium thermodynamical properties for the
corresponding system sizes.

\subsection{Grand-Canonical Annealing}

\label{sub:GCannealing} In parallel to the equilibrium properties
provided by Wang-Landau sampling, we also implement a form of simulated
annealing~\cite{SA} to explore the dynamics of the model and the
possibility of jamming, in a process that is similar to an experiment.
Simulated annealing used in this study follows the standard procedure
~\cite{Vcerny1985}. The corresponding experiment is exchanging particles
between the network and a reservoir of particles with (dimensionless)
chemical potential $\mu$. In our study, the annealing speed is not
controlled by decreasing temperature (which we set to $\beta=1$)
but by increasing the chemical potential. The annealing algorithm
is: 
\begin{enumerate}
\item Initially, start with chemical potential $\mu_{0}=0$ ; 
\item Randomly pick a site $n$; if it is empty (occupied), add (remove)
a particle with a probability of $\min\left[1,\exp(\mu)\right]$ ($\min\left[1,\exp(-\mu)\right]$)
while obeying the rule of the hard local density constraint on having
at most $l$ occupied nearest neighbors of $n$; 
\item If hopping is allowed, randomly pick one site; only if it is occupied,
randomly pick one of its empty neighbor(s) and displace the particle
if the density constraint remains satisfied; 
\item Increase $\mu$ by $d\mu$ every 1 Monte Carlo sweep ($N$ random
updates), where $d\mu/dt$ (in time-units of $dt=1$) is the annealing
schedule and $d\mu\ll1$; 
\item Repeat steps 2 to 4 until $\mu$ reaches a certain (large) chemical
potential. 
\end{enumerate}
Following the procedure above, the simulated annealing can reveal
whether or not a jamming transition occurs in the process. Besides
that, we can test the effect of local dynamics~\cite{Biroli2002,Krzakala2008}
by adding a local hopping random walk (step 3), i.e., a particle can
transfer any of its empty neighboring sites as long as 
the constraint remains satisfied.  The results are shown and explained in
the following section.

\section{Results}

\label{sec:results} To assess the properties of jamming, we first
have to benchmark our systems with the corresponding equilibrium behaviors.
After that, we discuss the dynamic simulations with the annealing
algorithm in reference to these equilibrium benchmarks.

\subsection{Equilibrium Properties\label{sub:Equilibrium-PropertiesWang-Landa}}

Wang-Landau sampling, as described in Sec.~\ref{sub:WLsampling},
is ideally suited for our purpose, since it provides access directly
to the density of states $g_{n}$ as a function of occupation number
$n$, which yields the partition function as
\begin{equation}
Z(\mu)=\sum_{n=0}^{n_{max}}g_{n}e^{n\mu}.\label{eq:Zmu}
\end{equation}
All thermodynamic quantities in the equilibrium can be obtained numerically
by summation of the formal derivates of $Z(\mu)$, such as those in
Eqs.~(\ref{eq:GCPdiff}), over all permissible occupation numbers
$0\leq n\leq n_{max}<N$. (For all $n_{max}<n\leq N$ it is $g_{n}=0$.) 

\begin{figure}
\centering \includegraphics[width=0.8\columnwidth]{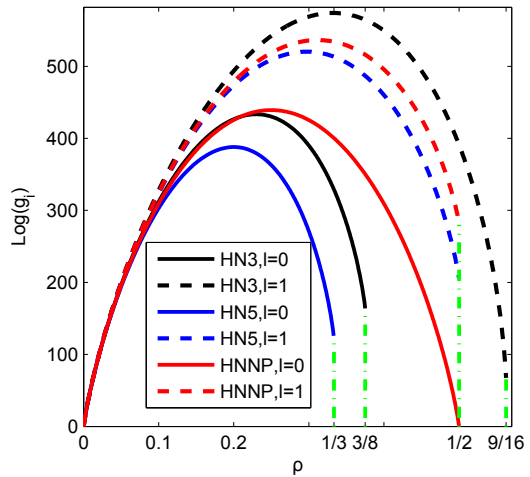}
\protect\caption{Density of states from Wang-Landau sampling at $N=1024$. The green
dash-dot vertical line are showing the closest packing fractions (as
shown in Table \ref{tab:cpf}) for each system. Note that only for
HNNP at $l=0$ there is a unique, crystalline ground state. }

\label{fig:doswl} 
\end{figure}

In Fig.~\ref{fig:doswl}, we plot the density of states as a function
of the packing fraction, both obtained with Wang-Landau. It becomes
apparent that each model has a simple rational value for its optimal
($\mu\to\infty$) ``random'' close packing fraction $\rho_{CP}=n_{max}/N$.
This corresponds to a random packing in the sense that it has a nontrivial
entropy density due to geometric disorder (imposed by the lack of
translational invariance in the lattice), except for HNNP at $l=0$,
which has a unique ``crystalline'' packing of every odd site being
occupied. While these values for $\rho_{CP}$ have been previously
obtained with RG for $l=0$~\cite{BoHa11}, the simulations predict
also strikingly simple but nontrivial values for $l=1$, where exact
RG is likely not possible. These values are listed in Table \ref{tab:cpf}.

\begin{table}
\begin{centering}
\protect\caption{\label{tab:cpf} Closest packing fractions $\rho_{CP}$ found by Wang-Landau sampling.
The values for $l=0$ have been previously obtained with exact RG,
the one for HNNP being unique, with every second, odd site occupied.
For $l=1$, we also predict exact fractions with nontrivial entropy
densities, see Fig.~\ref{fig:doswl}. }

\par\end{centering}

\begin{centering}

\par\end{centering}

\centering{}%
\begin{tabular}{|c|c||c|}
\hline 
Network & $l=0$  & $l=1$ \tabularnewline
\hline 
\hline 
HN3  & 3/8~\cite{BoHa11} & 9/16 \tabularnewline
\hline 
HN5  & 1/3~\cite{BoHa11} & 1/2 \tabularnewline
\hline 
HNNP  & 1/2  & 1/2 \tabularnewline
\hline 
\end{tabular}
\end{table}

\begin{figure}
\centering \includegraphics[width=1\columnwidth]{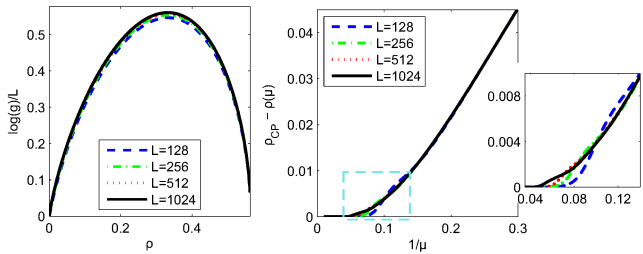}
\protect\caption{Convergence to the thermodynamic limit for finite system sizes for
the example of HN3 with $l=1$ using Wang-Landau sampling. The figures are
for the density of states (left) and the packing fraction (right). The
equilibrium packing fraction $\rho(\mu)$ as a function of chemical
potential $\mu$ is calculated from the density of states according
to Eq.~(\ref{eq:GCPdiff}); it approaches the closest packing fraction
$\rho_{CP}$ for $1/\mu\to0$. The convergence for other systems is
similar or better. }

\label{fig:WLconverge} 
\end{figure}

\begin{figure}
\includegraphics[width=0.8\columnwidth]{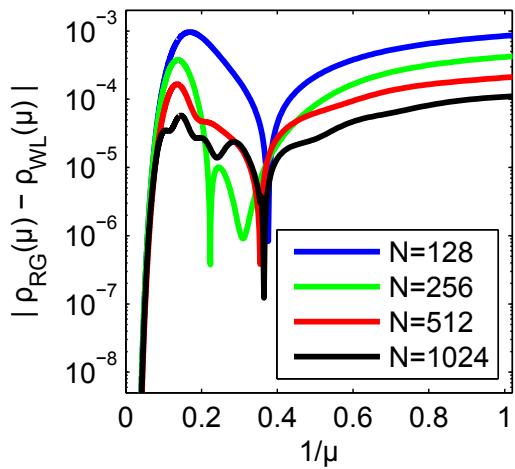}\protect\caption{\label{fig:RGvsWL}Plot of the error in the finite-size packing fraction
in Wang-Landau sampling, $\left|\rho_{WL}-\rho_{RG}\right|$, as a
function of $1/\mu$ near close packing ($\mu\to\infty$) in HN3 at
$l=0$. Here, RG result $\rho_{RG}(\mu)$ from Ref.~~\cite{BoHa11}
at system size $N=2^{500}$ is taken as the exact, thermodynamic packing
fraction. Relative to $\rho_{RG}(\mu)$, the finite-size packing fraction,
$\rho_{WL}(\mu)$, at $N=2^{k}$ with $k=8,9,10$ already exhibit
quite small and rapidly diminishing corrections. }
\end{figure}

Wang-Landau sampling converges within a reasonable time for system
sizes smaller than $N\approx2000$ but fails to converge for larger
system size. There may be two reasons for the lack of convergence:
(1) the density of states is not symmetric as a function of packing
fraction, and this asymmetry requires Wang-Landau to sample the whole
configuration space, which increases the computational cost dramatically
especially for large system sizes; (2) the lower the density of states
of the closest packed state, the harder it is for Monte Carlo sampling
to find its closest packing state because of the hard-density constraint.
Although Wang-Landau sampling fails for large system sizes, the results
of system size $N=1024$ can still offer an insight to the equilibrium
state because the density of states and the packing fraction exhibit only small 
finite-size corrections for increasing $N$. For example, the convergence of HN3
with $l=0$ is shown in Fig.\ref{fig:WLconverge}. Other networks
with $l=0,1$ have similar or even better convergence.

We can further demonstrate the quality of the Wang-Landau simulations,
and appraise their residual finite-size effects, by comparison with
exact results obtained with the renormalization group (RG) for $l=0$
on HN3~\cite{BoHa11}. In Fig.~\ref{fig:RGvsWL}, we compare the results
for the packing fraction $\rho(\mu)$ as a function of the chemical
potential for Wang-Landau sampling on networks with $N=2^{k}$ sites, $k=8\sim10$,
with those from the exact RG after 500 iterations, corresponding
to a system of $N=2^{500}$ sites. Despite the much smaller sizes
of the Wang-Landau simulation, its results are barely distinguishable
from the exact result, affirming the Wang-Landau sampling results
as good references for our dynamic simulations, with negligible finite-size
effects.

\subsection{Dynamic Properties}

\label{subsec:jamsc} The dynamic simulations of the BM on our networks
uses the grand canonical partition function controlled by a chemical
potential $\mu$ that mimics the experimental situation in a complex
fluid or colloid, where particles are pumped into the larger system
(the reservoir) and can enter the field-of-view through open boundaries
inside a smaller window. For example, this could correspond to a $2d$
slice of a $3d$ colloidal bath used in colloidal tracking experiments
~\cite{Hunter12}. Since our particles are not energetically coupled
and merely obey hard excluded volume constraints, temperature is irrelevant
and we can set $\beta=1$, making the chemical potential dimensionless,
$\beta\mu\to\mu$. As we increase $\mu$, the system is more likely
to accept more particles and increase the packing fraction $\rho(\mu)$.
When $\mu$ is small (or negative), the reservoir and the network
readily reach an equilibrium state with a certain packing fraction.
However, when $\mu$ is large, the equilibrium state defined by the
partition function has a packing fraction close to the close packing
$\rho_{CP}$. Because of the density constraint and the disorder imposed
by the hierarchical network geometry, the system enters into a jam
at a density far from equilibrium packing. As in experiments, this
jammed state remains for an extremely long time, even when $\mu$
is further increased. The ultimate packing fraction $\rho^{*}$ that
the systems gets stuck at, in fact, is ever further from random close
packing, the faster the quench in $\mu$ is executed, where $\frac{d\mu}{dt}$
is the quench rate. In this, our results closely resemble those reported
in Ref.~\cite{Krzakala2008}. 

\begin{figure}
\centering \includegraphics[width=1\columnwidth]{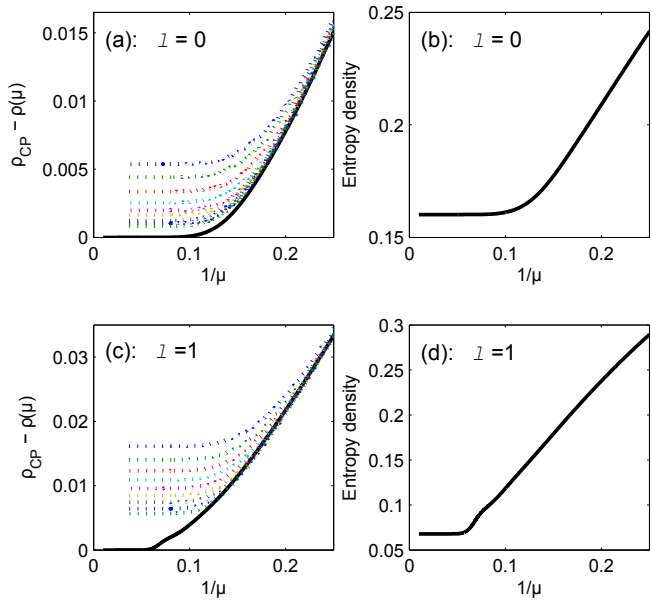}
\protect\caption{Reduced packing fraction and entropy density for HN3 from Wang-Landau
sampling and Simulated Annealing. (a)\&(b) are for $l=0$, and (c)
\&(d) are for $l=1$. The black solid lines represent the equilibrium
properties from Wang-Landau sampling with $N=1024$. The dotted lines
are from simulated annealing with $N=32,768$, run at different annealing
schedules with $d\mu=0.001/2^{j}$ for $j=0,\ldots,8$, from top to
bottom. Wang-Landau sampling provides the entropy density via Eq.
(\ref{eq:GCPdiff}), as shown in (b) and (d), which is difficult to
obtain from other Monte Carlo methods. For both, $l=0$ and 1, we
find a non-zero entropy density for random close packing at $\mu\rightarrow\infty$.}

\label{fig:HN3PE} 
\end{figure}

\begin{figure}
\centering \includegraphics[width=0.9\columnwidth]{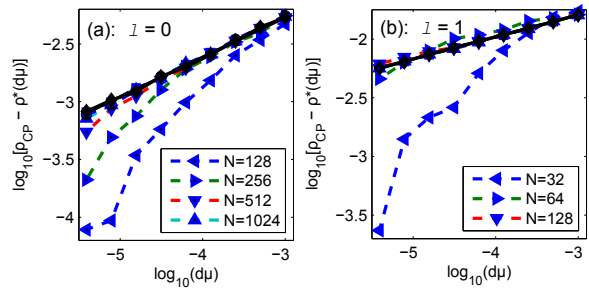}
\protect\caption{Scaling of the dynamically reached packing fraction $\rho^{*}(d\mu)$
as a function of the annealing rate $d\mu$ for different system sizes
$N$ of HN3. (a) For $l=0$, the dashed lines are for systems sizes
$N=2^{k}$ with $k=7,\ldots,$10, 12, 14 and 15, from bottom to top.
All data sets (except for the smallest sizes, $N=128,\ldots,1024$)
collapse onto the top line with a slope of $0.34\pm0.01$, which is
obtained from a fit using the data of the largest system size $N=32,768$.
(b) For $l=1$, the data sets converge even faster towards power-law
scaling. The dashed lines are for system sizes of $N=2^{k}$ with
$k=5,\ldots,8$, 10, 12, 14 and 15, from bottom to top. All but the
first 3 sets collapse onto a line of slope $0.19\pm0.01,$ which is
obtained from a fit for $N=32,768$. Error bars are about of the size
of each data point or smaller, indicating a relative error of less
than $3\%$.}

\label{fig:HN3Decay} 
\end{figure}

\begin{figure}
\centering \includegraphics[width=0.9\columnwidth]{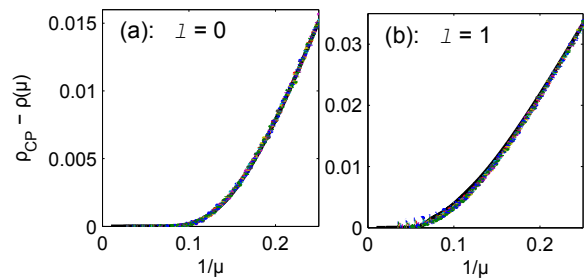}
\protect\caption{Results of simulated annealing with hopping for HN3. For both (a)
and (b), the figure consists of one solid line for the equilibrium
result obtained with Wang-Landau sampling and 9 dotted lines obtained
with simulated annealing at rates $d\mu=0.001/2^{j}$ for $j=0,\ldots,8$.
For HN3 with $l=0$ and $l=1$, the system equilibrates for nearly
all annealing schedules, collapsing the data onto the equilibrium
line. Only for HN3 with $l=1$, small deviations from equilibrium
are observed for annealing schedules $d\mu\gtrsim10^{-5}$. }

\label{fig:HN3hopping} 
\end{figure}

\subsubsection{Results for HN3}

\label{subsec:HN3}

The equilibrium packing fraction and entropy from Wang-Landau sampling
as well as the dynamic results from simulated annealing for HN3 are
shown in Fig.~\ref{fig:HN3PE}. Based on the analytical results by
Boettcher \textit{et al.}~\cite{BoHa11}, we can confidently conclude
that there is no phase transition in HN3 with $l=0$. Yet, the dynamic
simulations indicate that the system jams nonetheless. The system
jams even further from equilibrium for the case of $l=1$. Here, RG
results have not been obtained so far and it is not clear whether
there is a thermodynamic phase transition. The equilibrium results
from Wang-Landau sampling (at $N=2^{10}$) seem to suggest a singularity
near $1/\mu\approx0.06$ where the entropy density jumps noticeably
and $\rho(\mu)\equiv\rho_{CP}$ for all larger $\mu$. Either RG or
results for bigger systems may be needed to confirm whether there
is phase transition or not. 

The possible jamming transitions for both $l=0$ and $1$, revealed
by the dynamic annealing simulations in Fig.~\ref{fig:HN3PE} (a)
and (c), are further supported by a power law decay of the residual
packing fractions, $\rho_{CP}-\rho^{*}(d\mu)$, as a function of the
annealing rate, $d\mu$. Here, we set the jammed packing fraction,
obtained at $\mu\to\infty$ after annealing at rate $d\mu$, as $\rho^{*}(d\mu)=\rho(\mu\to\infty;d\mu)$,
where $d\mu/dt\to d\mu$ when measured in units of $dt\hat{=}1$ sweep.
Note that at these system sizes ($N=32,768$), even the weakest jam
is of order $\rho_{CP}-\rho^{*}(d\mu)\approx0.001$ and, thus, still
consists of a sizable number ($\gtrsim30$) of frustrated particles. 

As shown in Fig.~\ref{fig:HN3Decay}, a linear fit of the data on
a double-logarithmic scale at the largest systems  is
nearly perfect, justifying the assumption that the time-scales $1/d\mu$
for the existence of the jam diverge asymptotically with a power law
for $\rho\to\rho_{CP}$. For HN3 at $l=0$, the slope is $0.34\pm0.01$
with coefficient of determination $R^{2}=0.9975$, while for $l=1$
the slope is $0.19\pm0.01$ with $R^{2}=0.9997$, in both cases indicating
a dramatic increase of time-scales. 

We also test the effect of introducing local hopping, implemented
as suggested in step 3 of the algorithm in Sec.~\ref{sub:GCannealing},
which has not been addressed in Refs.~\cite{Krzakala2008, Biroli2002}.
The results shown in Fig.~\ref{fig:HN3hopping} indicate a substantial
difference from the simulation without hopping. For HN3 with $l=0$,
the jamming transition disappears even for the fastest annealing schedule,
$d\mu=10^{-3}$. For HN3 with $l=1$, the jamming transition can be
eliminated at least for an annealing schedule of $d\mu\approx10^{-5}$
or slower. 

Besides the Hanoi networks, we have repeated the annealing simulations on random regular graphs, 
following Krzakala {\it et al.}~\cite{Krzakala2008}. On those graphs, BM with a hopping dynamics 
can reach a much denser state than with a varying chemical potential alone, which is similar to what Rivoire {\it  et al.}~\cite{Rivoire03} argue. But because of the enormous computational cost, we can only test $d\mu$ to as small as $\sim10^{-6}$ for system sizes at most as large as $\sim 10^5$. No results are obtained to conclude that the jamming transition disappears entirely for some smaller $d\mu$, and we suspect that the behavior instead may resemble the mean-field predictions of Rivoire {\it et al.} \cite{Rivoire03}.  

\begin{figure}
\centering \includegraphics[width=1\columnwidth]{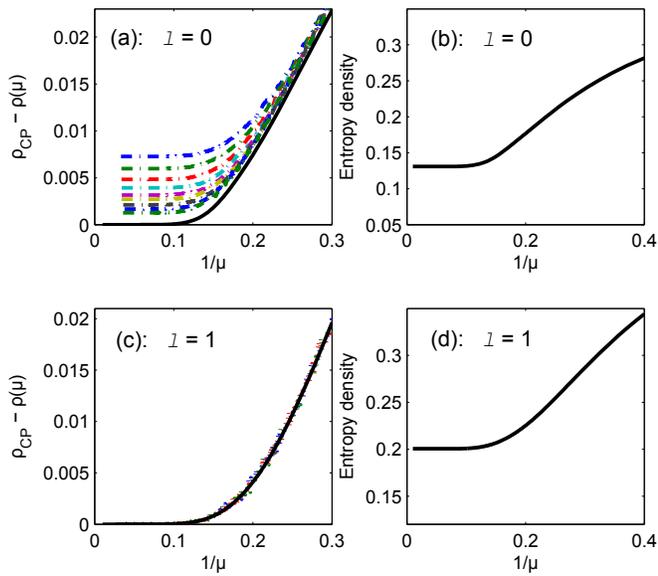}
\protect\caption{Reduced packing fraction and entropy density for HN5 from Wang-Landau
sampling and Simulated Annealing. (a)\&(b) are for $l=0$, and (c)
\&(d) are for $l=1$. The black solid lines represent the equilibrium
properties from Wang-Landau sampling with $N=1024$. The dotted lines
are from simulated annealing with $N=32,768$, run at different annealing
schedules with $d\mu=0.001/2^{j}$ for $j=0,\ldots,8$, from top to
bottom. As in Fig.~\ref{fig:HN3PE}, Wang-Landau sampling provides
the entropy density via Eq.~(\ref{eq:GCPdiff}), as shown in (b) and
(d).}
\label{fig:HN5PE} 
\end{figure}

\begin{figure}
\centering \includegraphics[width=0.8\columnwidth]{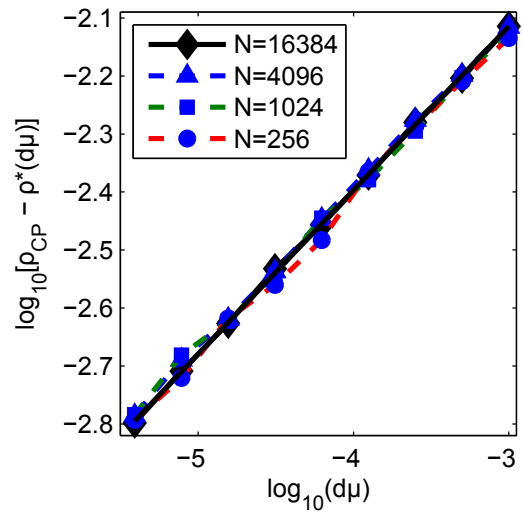}
\protect\caption{Scaling of the dynamically reached packing fraction $\rho^{*}(d\mu)$
as a function of the annealing rate $d\mu$ for different system sizes
$N$ of HN5 for $l=0$, the dashed lines are for systems sizes $N=2^{k}$
with $k=8,10,12$, and 14. All data sets collapse onto the black solid line
with a slope of $0.31\pm0.01$ with $R^{2}=0.9989$, which is obtained
from a fit using the data of the largest system size $N=16,384$. Error
bars are about of the size of each data point or smaller, indicating
a relative error of less than $3\%$. }

\label{fig:HN5Decay} 
\end{figure}

\subsubsection{Results for HN5}

\label{subsec:HN5}

The case in HN5 is different from that in HN3. Note that HN5, unlike
HN3 and most finite-dimensional lattices or the random graphs studied
in Ref.~~\cite{Krzakala2008}, is not a regular network but has an exponential
degree distribution. In HN5 for both, $l=0$ and $l=1$, as shown
in Fig.~\ref{fig:HN5PE}, the equilibrium behavior obtained from Wang-Landau
sampling is smooth and there is no indication of a phase transition.
Annealing reveals a jamming transition and a power law decay similar
to that in HN3 in the dynamic simulations only for $l=0$. For $l=1$,
surprisingly, there is no jamming transition. The simulations with
different annealing schedules equilibrate easily and collapse with
the curves from Wang-Landau sampling. This suggests that the combination
of heterogeneity in neighborhood sizes together with the possibility
to have one occupied neighbor ``lubricates'' the system sufficiently
to avert jams. Correspondingly, the results from Wang-Landau converge
rapidly even for larger system sizes. As for HN3, permitting a local
hopping dynamics unjams the system also for HN5 with $l=0$.

\begin{figure}
\centering \includegraphics[width=0.95\columnwidth]{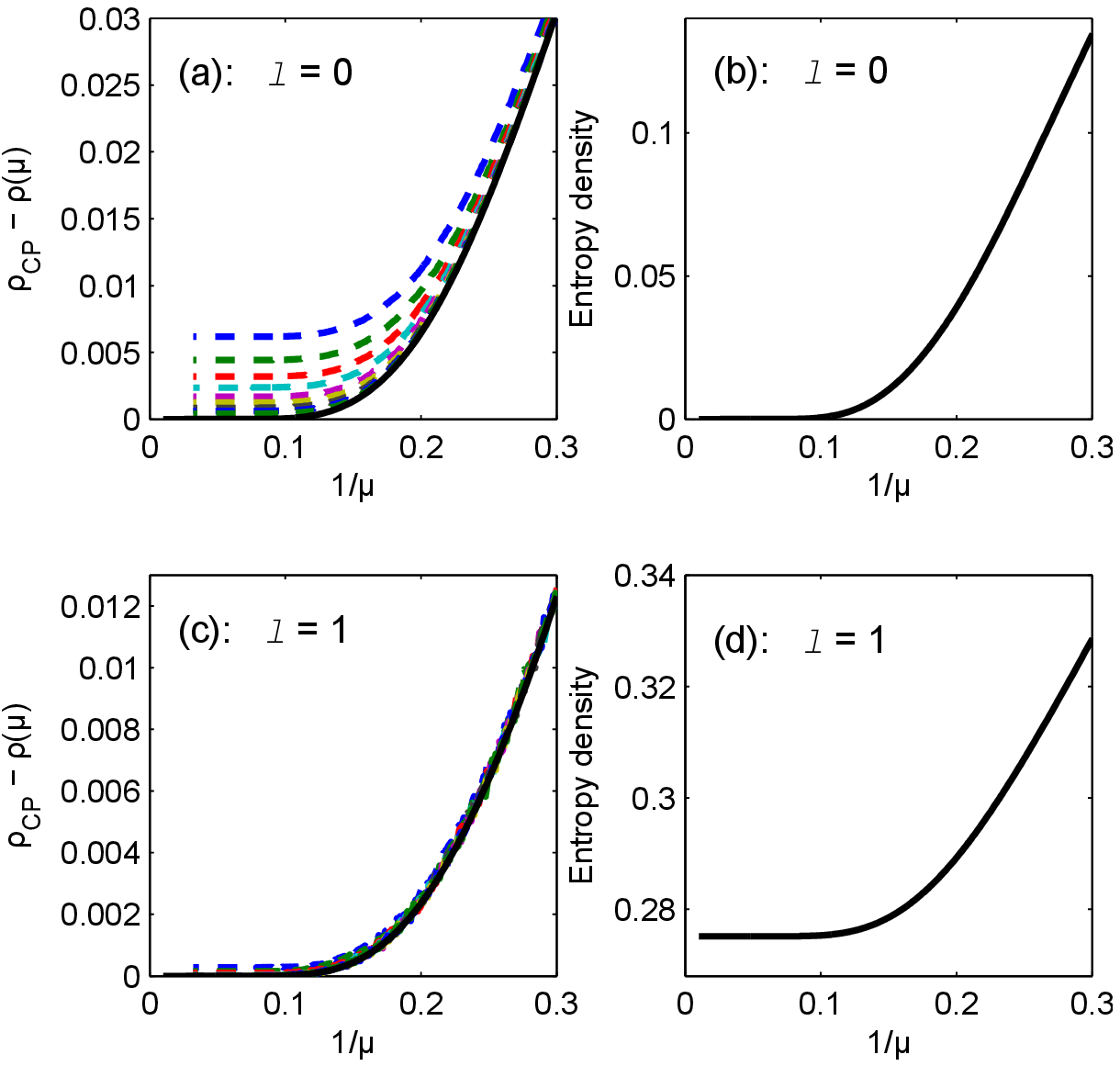}
\protect\caption{Reduced packing fraction and entropy density for HNNP from Wang-Landau
sampling and Simulated Annealing. (a)\&(b) are for $l=0$, and (c)
\&(d) are for $l=1$. The black solid lines represent the equilibrium
properties from Wang-Landau sampling with $N=1024$. The dotted lines
are from simulated annealing with $N=16,384$, run at different annealing
schedules with $d\mu=0.001/2^{j}$ for $j=0,\ldots,8$, from top to
bottom. As in Figs. \ref{fig:HN3PE} and \ref{fig:HN5PE}, Wang-Landau
sampling provides the entropy density via Eq.~(\ref{eq:GCPdiff}),
as shown in (b) and (d). Note that in the limit of $\mu\to\infty$,
HNNP at $l=0$ has a zero entropy which corresponds to a unique ground
state. At $l=1$, it attains the \emph{same} close packing fraction,
$\rho_{CP}=\frac{1}{2}$, see Table \ref{tab:cpf}, but now at a non-trivial
entropy. }

\label{fig:HNNPPE} 
\end{figure}

\begin{figure}
\centering \includegraphics[width=0.8\columnwidth]{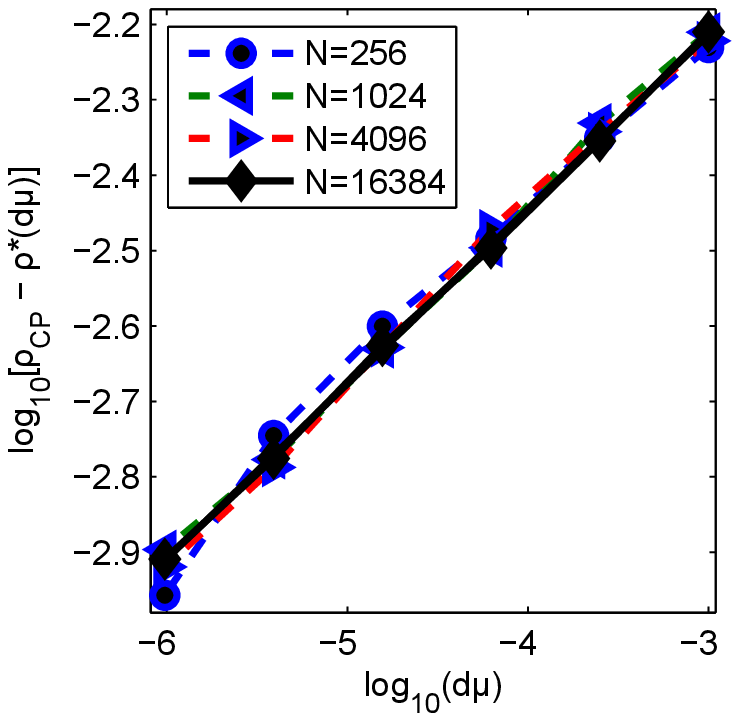}
\protect\caption{Scaling of the dynamically reached packing fraction $\rho^{*}(d\mu)$
as a function of the annealing rate $d\mu$ for different system sizes
$N$ of HNNP for $l=0$, the dashed lines are for systems sizes $N=2^{k}$
with $k=8,10,12$, and 14. All data sets collapse onto the top line
with a slope of $0.23\pm0.01$ with $R^{2}=0.9997$, which is obtained
from a fit using the data of the largest system size $N=16,384$. Error
bars are about of the size of each data point or smaller, indicating
a relative error of less than $3\%$. }

\label{fig:HNNPDecay} 
\end{figure}

\begin{table*}
\begin{centering}
\protect\caption{Summary of the results. For each network, and the allowed neighborhood
occupations of $l=0$ and $l=1$, we list to potential for a jam in
dynamic simulations and the likely existence of an equilibrium glass
transition.}

\par\end{centering}

\begin{centering}
\label{tab:summary}
\par\end{centering}

\centering{}%
\begin{tabular}{|c|c||c|}
\hline 
 & $l=0$  & $l=1$ \tabularnewline
\hline 
\hline 
HN3  & Jamming transition \& no phase transition  & Jamming transition \& uncertain \tabularnewline
\hline 
HN5  & Jamming transition \& uncertain  & No jamming transition \& no phase transition \tabularnewline
\hline 
HNNP  & Jamming transition \& uncertain  & No jamming transition\& no phase transition \tabularnewline
\hline 
\end{tabular}
\end{table*}

\subsubsection{Results for HNNP}

\label{subsec:HNNP}

HNNP provides an interesting alternative among the networks we are
considering here. Unlike HN3 and HN5, HNNP is a nonplanar network,
but like HN5 it has an exponential distribution of degrees with an
average degree of 4. Most importantly, HNNP at $l=0$ possesses a
``crystalline'' optimal packing that is unique, see Fig.~\ref{fig:HNNPPE}(b),
and consists of every second site along the line being occupied, i.e.,
those sites that uniformly have the lowest degree of 3. Therefore,
it provides the opportunity to explore the potential for a first-order
transition from a jammed state into the ground state, as was observed
for lattice glasses in Ref.~~\cite{Biroli02}. In this case, RG can
be applied to obtain $\rho(\mu)$ in equilibrium exactly.

Indeed, we find a weakly jammed state in HNNP with $l=0$, with only
a small number of frustrated particles, as shown in Fig.~\ref{fig:HNNPPE}.
The results of annealing simulations also show a power-law decay (Fig
\ref{fig:HNNPDecay}), consistent with the approach to a jamming transition.
As RG suggest, and the smooth equilibrium curve for $N=1024$ and
the convergence with increasing system sizes affirm, there is no thermodynamic
phase transition in HNNP with $l=0$. Despite the weakness of those
jams, we can find no indication that the annealing simulations at
any rate $d\mu$ can ever decay into the ordered state. Apparently,
the structural disorder, enforced in HNNP through a heterogeneous
neighborhood degree and the hierarchy of long-range links, prevents
such an explosive transition. The dominance of such structural elements
is further emphasized by the fact that HNNP for $l=1$ exhibits no
jams, similar to HN5, with which HNNP shares that structure.

\section{Conclusions}

\label{sec:conclusions}

We have examined the Biroli-Mezard lattice glass model on hierarchical
networks, which provide intermediaries between solvable mean-field
models and intractable finite-dimensional systems. These networks
exhibit a lattice-like structure with small loops but also with a hierarchy of long-range links
imposing geometric disorder and frustration while preserving a recursive
structure that can be explored with exact methods, in principle. We
observed a rich variety of dynamic behaviors in our simulations. For
instance, we find jamming behavior on a regular network for which
RG has shown that no equilibrium phase transition exists. However,
whether the dynamic transition occurs at a packing fraction distinctly
above random close packing remains unclear, and can only be resolved
with more detailed RG studies that are beyond our discussion here.

We have simulated the model on our networks with a varying chemical potential $\mu$,
with and without local hopping of particles. Hopping impacted those
simulations in a significant manner, always eliminating any jams that have existed without hopping. Solutions of the corresponding mean-field systems  would have suggested that a dynamics driven by hopping (but at fixed particle number) results in kinetic arrest~\cite{Rivoire03}. Whether canonical simulations with
hopping alone, or hopping at different rates, would change this scenario, we have to leave for future investigations, as well as the question
on whether a combined method of updates would alter the behavior observed
on lattices and mean-field networks.

\section*{Acknowledgements}

We thank the U. S. National Science Foundation for its support through
grant DMR-1207431.
\bibliographystyle{apsrev4-1}
\bibliography{Boettcher}

\begin{thebibliography}{29}%
\makeatletter
\providecommand \@ifxundefined [1]{%
 \@ifx{#1\undefined}
}%
\providecommand \@ifnum [1]{%
 \ifnum #1\expandafter \@firstoftwo
 \else \expandafter \@secondoftwo
 \fi
}%
\providecommand \@ifx [1]{%
 \ifx #1\expandafter \@firstoftwo
 \else \expandafter \@secondoftwo
 \fi
}%
\providecommand \natexlab [1]{#1}%
\providecommand \enquote  [1]{``#1''}%
\providecommand \bibnamefont  [1]{#1}%
\providecommand \bibfnamefont [1]{#1}%
\providecommand \citenamefont [1]{#1}%
\providecommand \href@noop [0]{\@secondoftwo}%
\providecommand \href [0]{\begingroup \@sanitize@url \@href}%
\providecommand \@href[1]{\@@startlink{#1}\@@href}%
\providecommand \@@href[1]{\endgroup#1\@@endlink}%
\providecommand \@sanitize@url [0]{\catcode `\\12\catcode `\$12\catcode
  `\&12\catcode `\#12\catcode `\^12\catcode `\_12\catcode `\%12\relax}%
\providecommand \@@startlink[1]{}%
\providecommand \@@endlink[0]{}%
\providecommand \url  [0]{\begingroup\@sanitize@url \@url }%
\providecommand \@url [1]{\endgroup\@href {#1}{\urlprefix }}%
\providecommand \urlprefix  [0]{URL }%
\providecommand \Eprint [0]{\href }%
\providecommand \doibase [0]{http://dx.doi.org/}%
\providecommand \selectlanguage [0]{\@gobble}%
\providecommand \bibinfo  [0]{\@secondoftwo}%
\providecommand \bibfield  [0]{\@secondoftwo}%
\providecommand \translation [1]{[#1]}%
\providecommand \BibitemOpen [0]{}%
\providecommand \bibitemStop [0]{}%
\providecommand \bibitemNoStop [0]{.\EOS\space}%
\providecommand \EOS [0]{\spacefactor3000\relax}%
\providecommand \BibitemShut  [1]{\csname bibitem#1\endcsname}%
\let\auto@bib@innerbib\@empty
\bibitem [{\citenamefont {Liu}\ and\ \citenamefont {Nagel}(1998)}]{Liu1998}%
  \BibitemOpen
  \bibfield  {author} {\bibinfo {author} {\bibfnamefont {A.~J.}\ \bibnamefont
  {Liu}}\ and\ \bibinfo {author} {\bibfnamefont {S.~R.}\ \bibnamefont
  {Nagel}},\ }\href@noop {} {\bibfield  {journal} {\bibinfo  {journal}
  {Nature}\ }\textbf {\bibinfo {volume} {396}},\ \bibinfo {pages} {21}
  (\bibinfo {year} {1998})}\BibitemShut {NoStop}%
\bibitem [{\citenamefont {Biroli}(2007)}]{Biroli2007}%
  \BibitemOpen
  \bibfield  {author} {\bibinfo {author} {\bibfnamefont {G.}~\bibnamefont
  {Biroli}},\ }\href@noop {} {\bibfield  {journal} {\bibinfo  {journal} {Nature
  Physics}\ }\textbf {\bibinfo {volume} {3}},\ \bibinfo {pages} {222} (\bibinfo
  {year} {2007})}\BibitemShut {NoStop}%
\bibitem [{\citenamefont {Liu}\ and\ \citenamefont {Nagel}(2010)}]{Liu2010}%
  \BibitemOpen
  \bibfield  {author} {\bibinfo {author} {\bibfnamefont {A.~J.}\ \bibnamefont
  {Liu}}\ and\ \bibinfo {author} {\bibfnamefont {S.~R.}\ \bibnamefont
  {Nagel}},\ }\href@noop {} {\bibfield  {journal} {\bibinfo  {journal} {Annu.
  Rev. Condens. Matter Phys.}\ }\textbf {\bibinfo {volume} {1}},\ \bibinfo
  {pages} {347} (\bibinfo {year} {2010})}\BibitemShut {NoStop}%
\bibitem [{\citenamefont {Hill}\ and\ \citenamefont
  {Dissado}(1985)}]{Hill1985}%
  \BibitemOpen
  \bibfield  {author} {\bibinfo {author} {\bibfnamefont {R.}~\bibnamefont
  {Hill}}\ and\ \bibinfo {author} {\bibfnamefont {L.}~\bibnamefont {Dissado}},\
  }\href@noop {} {\bibfield  {journal} {\bibinfo  {journal} {Journal of Physics
  C: Solid State Physics}\ }\textbf {\bibinfo {volume} {18}},\ \bibinfo {pages}
  {3829} (\bibinfo {year} {1985})}\BibitemShut {NoStop}%
\bibitem [{\citenamefont {Ciamarra}\ \emph {et~al.}(2010)\citenamefont
  {Ciamarra}, \citenamefont {Nicodemi},\ and\ \citenamefont
  {Coniglio}}]{Ciamarra2010}%
  \BibitemOpen
  \bibfield  {author} {\bibinfo {author} {\bibfnamefont {M.~P.}\ \bibnamefont
  {Ciamarra}}, \bibinfo {author} {\bibfnamefont {M.}~\bibnamefont {Nicodemi}},
  \ and\ \bibinfo {author} {\bibfnamefont {A.}~\bibnamefont {Coniglio}},\
  }\href@noop {} {\bibfield  {journal} {\bibinfo  {journal} {Soft Matter}\
  }\textbf {\bibinfo {volume} {6}},\ \bibinfo {pages} {2871} (\bibinfo {year}
  {2010})}\BibitemShut {NoStop}%
\bibitem [{\citenamefont {Van~Hecke}(2010)}]{van2010}%
  \BibitemOpen
  \bibfield  {author} {\bibinfo {author} {\bibfnamefont {M.}~\bibnamefont
  {Van~Hecke}},\ }\href@noop {} {\bibfield  {journal} {\bibinfo  {journal}
  {Journal of Physics: Condensed Matter}\ }\textbf {\bibinfo {volume} {22}},\
  \bibinfo {pages} {033101} (\bibinfo {year} {2010})}\BibitemShut {NoStop}%
\bibitem [{\citenamefont {Majmudar}\ \emph {et~al.}(2007)\citenamefont
  {Majmudar}, \citenamefont {Sperl}, \citenamefont {Luding},\ and\
  \citenamefont {Behringer}}]{Majmudar2007}%
  \BibitemOpen
  \bibfield  {author} {\bibinfo {author} {\bibfnamefont {T.~S.}\ \bibnamefont
  {Majmudar}}, \bibinfo {author} {\bibfnamefont {M.}~\bibnamefont {Sperl}},
  \bibinfo {author} {\bibfnamefont {S.}~\bibnamefont {Luding}}, \ and\ \bibinfo
  {author} {\bibfnamefont {R.~P.}\ \bibnamefont {Behringer}},\ }\href {\doibase
  10.1103/PhysRevLett.98.058001} {\bibfield  {journal} {\bibinfo  {journal}
  {Phys. Rev. Lett.}\ }\textbf {\bibinfo {volume} {98}},\ \bibinfo {pages}
  {058001} (\bibinfo {year} {2007})}\BibitemShut {NoStop}%
\bibitem [{\citenamefont {Parisi}\ and\ \citenamefont
  {Zamponi}(2010)}]{Parisi2010}%
  \BibitemOpen
  \bibfield  {author} {\bibinfo {author} {\bibfnamefont {G.}~\bibnamefont
  {Parisi}}\ and\ \bibinfo {author} {\bibfnamefont {F.}~\bibnamefont
  {Zamponi}},\ }\href {\doibase 10.1103/RevModPhys.82.789} {\bibfield
  {journal} {\bibinfo  {journal} {Rev. Mod. Phys.}\ }\textbf {\bibinfo {volume}
  {82}},\ \bibinfo {pages} {789} (\bibinfo {year} {2010})}\BibitemShut
  {NoStop}%
\bibitem [{\citenamefont {Angelani}\ and\ \citenamefont
  {Foffi}(2007)}]{Angelani2007}%
  \BibitemOpen
  \bibfield  {author} {\bibinfo {author} {\bibfnamefont {L.}~\bibnamefont
  {Angelani}}\ and\ \bibinfo {author} {\bibfnamefont {G.}~\bibnamefont
  {Foffi}},\ }\href@noop {} {\bibfield  {journal} {\bibinfo  {journal} {Journal
  of Physics: Condensed Matter}\ }\textbf {\bibinfo {volume} {19}},\ \bibinfo
  {pages} {256207} (\bibinfo {year} {2007})}\BibitemShut {NoStop}%
\bibitem [{\citenamefont {Trappe}\ \emph {et~al.}(2001)\citenamefont {Trappe},
  \citenamefont {Prasad}, \citenamefont {Cipelletti}, \citenamefont {Segre},\
  and\ \citenamefont {Weitz}}]{Trappe2001}%
  \BibitemOpen
  \bibfield  {author} {\bibinfo {author} {\bibfnamefont {V.}~\bibnamefont
  {Trappe}}, \bibinfo {author} {\bibfnamefont {V.}~\bibnamefont {Prasad}},
  \bibinfo {author} {\bibfnamefont {L.}~\bibnamefont {Cipelletti}}, \bibinfo
  {author} {\bibfnamefont {P.}~\bibnamefont {Segre}}, \ and\ \bibinfo {author}
  {\bibfnamefont {D.}~\bibnamefont {Weitz}},\ }\href@noop {} {\bibfield
  {journal} {\bibinfo  {journal} {Nature}\ }\textbf {\bibinfo {volume} {411}},\
  \bibinfo {pages} {772} (\bibinfo {year} {2001})}\BibitemShut {NoStop}%
\bibitem [{\citenamefont {Zhang}\ and\ \citenamefont
  {Makse}(2005)}]{Zhang2005}%
  \BibitemOpen
  \bibfield  {author} {\bibinfo {author} {\bibfnamefont {H.}~\bibnamefont
  {Zhang}}\ and\ \bibinfo {author} {\bibfnamefont {H.}~\bibnamefont {Makse}},\
  }\href@noop {} {\bibfield  {journal} {\bibinfo  {journal} {Physical Review
  E}\ }\textbf {\bibinfo {volume} {72}},\ \bibinfo {pages} {011301} (\bibinfo
  {year} {2005})}\BibitemShut {NoStop}%
\bibitem [{\citenamefont {Berthier}\ \emph {et~al.}(2011)\citenamefont
  {Berthier}, \citenamefont {Biroli}, \citenamefont {Bouchaud}, \citenamefont
  {Cipelletti},\ and\ \citenamefont {van Saarloos}}]{Berthier2011}%
  \BibitemOpen
  \bibfield  {author} {\bibinfo {author} {\bibfnamefont {L.}~\bibnamefont
  {Berthier}}, \bibinfo {author} {\bibfnamefont {G.}~\bibnamefont {Biroli}},
  \bibinfo {author} {\bibfnamefont {J.-P.}\ \bibnamefont {Bouchaud}}, \bibinfo
  {author} {\bibfnamefont {L.}~\bibnamefont {Cipelletti}}, \ and\ \bibinfo
  {author} {\bibfnamefont {W.}~\bibnamefont {van Saarloos}},\ }\href@noop {}
  {\emph {\bibinfo {title} {Dynamical heterogeneities in glasses, colloids, and
  granular media}}}\ (\bibinfo  {publisher} {Oxford University Press},\
  \bibinfo {year} {2011})\BibitemShut {NoStop}%
\bibitem [{\citenamefont {Da~Cruz}\ \emph {et~al.}(2002)\citenamefont
  {Da~Cruz}, \citenamefont {Chevoir}, \citenamefont {Bonn},\ and\ \citenamefont
  {Coussot}}]{DaCruz2002}%
  \BibitemOpen
  \bibfield  {author} {\bibinfo {author} {\bibfnamefont {F.}~\bibnamefont
  {Da~Cruz}}, \bibinfo {author} {\bibfnamefont {F.}~\bibnamefont {Chevoir}},
  \bibinfo {author} {\bibfnamefont {D.}~\bibnamefont {Bonn}}, \ and\ \bibinfo
  {author} {\bibfnamefont {P.}~\bibnamefont {Coussot}},\ }\href@noop {}
  {\bibfield  {journal} {\bibinfo  {journal} {Physical Review E}\ }\textbf
  {\bibinfo {volume} {66}},\ \bibinfo {pages} {051305} (\bibinfo {year}
  {2002})}\BibitemShut {NoStop}%
\bibitem [{\citenamefont {Krzakala}\ \emph {et~al.}(2008)\citenamefont
  {Krzakala}, \citenamefont {Tarzia},\ and\ \citenamefont
  {Zdeborov\'a}}]{Krzakala2008}%
  \BibitemOpen
  \bibfield  {author} {\bibinfo {author} {\bibfnamefont {F.}~\bibnamefont
  {Krzakala}}, \bibinfo {author} {\bibfnamefont {M.}~\bibnamefont {Tarzia}}, \
  and\ \bibinfo {author} {\bibfnamefont {L.}~\bibnamefont {Zdeborov\'a}},\
  }\href {\doibase 10.1103/PhysRevLett.101.165702} {\bibfield  {journal}
  {\bibinfo  {journal} {Phys. Rev. Lett.}\ }\textbf {\bibinfo {volume} {101}},\
  \bibinfo {pages} {165702} (\bibinfo {year} {2008})}\BibitemShut {NoStop}%
\bibitem [{\citenamefont {Jacquin}\ \emph {et~al.}(2011)\citenamefont
  {Jacquin}, \citenamefont {Berthier},\ and\ \citenamefont
  {Zamponi}}]{Jacquin2011}%
  \BibitemOpen
  \bibfield  {author} {\bibinfo {author} {\bibfnamefont {H.}~\bibnamefont
  {Jacquin}}, \bibinfo {author} {\bibfnamefont {L.}~\bibnamefont {Berthier}}, \
  and\ \bibinfo {author} {\bibfnamefont {F.}~\bibnamefont {Zamponi}},\ }\href
  {\doibase 10.1103/PhysRevLett.106.135702} {\bibfield  {journal} {\bibinfo
  {journal} {Phys. Rev. Lett.}\ }\textbf {\bibinfo {volume} {106}},\ \bibinfo
  {pages} {135702} (\bibinfo {year} {2011})}\BibitemShut {NoStop}%
\bibitem [{\citenamefont {Biroli}\ and\ \citenamefont
  {Mezard}(2002)}]{Biroli02}%
  \BibitemOpen
  \bibfield  {author} {\bibinfo {author} {\bibfnamefont {G.}~\bibnamefont
  {Biroli}}\ and\ \bibinfo {author} {\bibfnamefont {M.}~\bibnamefont
  {Mezard}},\ }\href@noop {} {\bibfield  {journal} {\bibinfo  {journal} {Phys.
  Rev. Lett.}\ }\textbf {\bibinfo {volume} {88}},\ \bibinfo {pages} {025501}
  (\bibinfo {year} {2002})}\BibitemShut {NoStop}%
\bibitem [{\citenamefont {Kob}\ and\ \citenamefont {Andersen}(1993)}]{Kob93}%
  \BibitemOpen
  \bibfield  {author} {\bibinfo {author} {\bibfnamefont {W.}~\bibnamefont
  {Kob}}\ and\ \bibinfo {author} {\bibfnamefont {H.~C.}\ \bibnamefont
  {Andersen}},\ }\href {\doibase 10.1103/PhysRevE.48.4364} {\bibfield
  {journal} {\bibinfo  {journal} {Phys. Rev. E}\ }\textbf {\bibinfo {volume}
  {48}},\ \bibinfo {pages} {4364} (\bibinfo {year} {1993})}\BibitemShut
  {NoStop}%
\bibitem [{\citenamefont {Rivoire}\ \emph {et~al.}(2003)\citenamefont
  {Rivoire}, \citenamefont {Biroli}, \citenamefont {Martin},\ and\
  \citenamefont {M{\'e}zard}}]{Rivoire03}%
  \BibitemOpen
  \bibfield  {author} {\bibinfo {author} {\bibfnamefont {O.}~\bibnamefont
  {Rivoire}}, \bibinfo {author} {\bibfnamefont {G.}~\bibnamefont {Biroli}},
  \bibinfo {author} {\bibfnamefont {O.~C.}\ \bibnamefont {Martin}}, \ and\
  \bibinfo {author} {\bibfnamefont {M.}~\bibnamefont {M{\'e}zard}},\
  }\href@noop {} {\bibfield  {journal} {\bibinfo  {journal} {The European
  Physical Journal B-Condensed Matter and Complex Systems}\ }\textbf {\bibinfo
  {volume} {37}},\ \bibinfo {pages} {55} (\bibinfo {year} {2003})}\BibitemShut
  {NoStop}%
\bibitem [{\citenamefont {Boettcher}\ \emph {et~al.}(2008)\citenamefont
  {Boettcher}, \citenamefont {Gon{\c{c}}alves},\ and\ \citenamefont
  {Guclu}}]{Boettcher2008HN}%
  \BibitemOpen
  \bibfield  {author} {\bibinfo {author} {\bibfnamefont {S.}~\bibnamefont
  {Boettcher}}, \bibinfo {author} {\bibfnamefont {B.}~\bibnamefont
  {Gon{\c{c}}alves}}, \ and\ \bibinfo {author} {\bibfnamefont {H.}~\bibnamefont
  {Guclu}},\ }\href@noop {} {\bibfield  {journal} {\bibinfo  {journal} {Journal
  of Physics A: Mathematical and Theoretical}\ }\textbf {\bibinfo {volume}
  {41}},\ \bibinfo {pages} {252001} (\bibinfo {year} {2008})}\BibitemShut
  {NoStop}%
\bibitem [{\citenamefont {Boettcher}\ and\ \citenamefont
  {Hartmann}(2011)}]{BoHa11}%
  \BibitemOpen
  \bibfield  {author} {\bibinfo {author} {\bibfnamefont {S.}~\bibnamefont
  {Boettcher}}\ and\ \bibinfo {author} {\bibfnamefont {A.~K.}\ \bibnamefont
  {Hartmann}},\ }\href@noop {} {\bibfield  {journal} {\bibinfo  {journal}
  {Phys. Rev. E}\ }\textbf {\bibinfo {volume} {84}},\ \bibinfo {pages} {011108}
  (\bibinfo {year} {2011})}\BibitemShut {NoStop}%
\bibitem [{\citenamefont {Wang}\ and\ \citenamefont
  {Landau}(2001{\natexlab{a}})}]{Wang2001}%
  \BibitemOpen
  \bibfield  {author} {\bibinfo {author} {\bibfnamefont {F.}~\bibnamefont
  {Wang}}\ and\ \bibinfo {author} {\bibfnamefont {D.~P.}\ \bibnamefont
  {Landau}},\ }\href {\doibase 10.1103/PhysRevLett.86.2050} {\bibfield
  {journal} {\bibinfo  {journal} {Phys. Rev. Lett.}\ }\textbf {\bibinfo
  {volume} {86}},\ \bibinfo {pages} {2050} (\bibinfo {year}
  {2001}{\natexlab{a}})}\BibitemShut {NoStop}%
\bibitem [{\citenamefont {Wang}\ and\ \citenamefont
  {Landau}(2001{\natexlab{b}})}]{wang:01a}%
  \BibitemOpen
  \bibfield  {author} {\bibinfo {author} {\bibfnamefont {F.}~\bibnamefont
  {Wang}}\ and\ \bibinfo {author} {\bibfnamefont {D.~P.}\ \bibnamefont
  {Landau}},\ }\href@noop {} {\bibfield  {journal} {\bibinfo  {journal} {Phys.
  Rev. E}\ }\textbf {\bibinfo {volume} {64}},\ \bibinfo {pages} {056101}
  (\bibinfo {year} {2001}{\natexlab{b}})}\BibitemShut {NoStop}%
\bibitem [{\citenamefont {Boettcher}\ \emph {et~al.}(2009)\citenamefont
  {Boettcher}, \citenamefont {Cook},\ and\ \citenamefont
  {Ziff}}]{Boettcher09c}%
  \BibitemOpen
  \bibfield  {author} {\bibinfo {author} {\bibfnamefont {S.}~\bibnamefont
  {Boettcher}}, \bibinfo {author} {\bibfnamefont {J.~L.}\ \bibnamefont {Cook}},
  \ and\ \bibinfo {author} {\bibfnamefont {R.~M.}\ \bibnamefont {Ziff}},\
  }\href@noop {} {\bibfield  {journal} {\bibinfo  {journal} {Phys. Rev. E}\
  }\textbf {\bibinfo {volume} {80}},\ \bibinfo {pages} {041115} (\bibinfo
  {year} {2009})}\BibitemShut {NoStop}%
\bibitem [{\citenamefont {Lee}\ \emph {et~al.}(2006)\citenamefont {Lee},
  \citenamefont {Okabe},\ and\ \citenamefont {Landau}}]{Lee2006}%
  \BibitemOpen
  \bibfield  {author} {\bibinfo {author} {\bibfnamefont {H.~K.}\ \bibnamefont
  {Lee}}, \bibinfo {author} {\bibfnamefont {Y.}~\bibnamefont {Okabe}}, \ and\
  \bibinfo {author} {\bibfnamefont {D.}~\bibnamefont {Landau}},\ }\href@noop {}
  {\bibfield  {journal} {\bibinfo  {journal} {Computer physics communications}\
  }\textbf {\bibinfo {volume} {175}},\ \bibinfo {pages} {36} (\bibinfo {year}
  {2006})}\BibitemShut {NoStop}%
\bibitem [{\citenamefont {Cunha-Netto}\ and\ \citenamefont
  {Dickman}(2011)}]{Dickman2011}%
  \BibitemOpen
  \bibfield  {author} {\bibinfo {author} {\bibfnamefont {A.}~\bibnamefont
  {Cunha-Netto}}\ and\ \bibinfo {author} {\bibfnamefont {R.}~\bibnamefont
  {Dickman}},\ }\href@noop {} {\bibfield  {journal} {\bibinfo  {journal}
  {Computer Physics Communications}\ }\textbf {\bibinfo {volume} {182}},\
  \bibinfo {pages} {719} (\bibinfo {year} {2011})}\BibitemShut {NoStop}%
\bibitem [{\citenamefont {Kirkpatrick}\ \emph {et~al.}(1983)\citenamefont
  {Kirkpatrick}, \citenamefont {Gelatt},\ and\ \citenamefont {Vecchi}}]{SA}%
  \BibitemOpen
  \bibfield  {author} {\bibinfo {author} {\bibfnamefont {S.}~\bibnamefont
  {Kirkpatrick}}, \bibinfo {author} {\bibfnamefont {C.~D.}\ \bibnamefont
  {Gelatt}}, \ and\ \bibinfo {author} {\bibfnamefont {M.~P.}\ \bibnamefont
  {Vecchi}},\ }\href@noop {} {\bibfield  {journal} {\bibinfo  {journal}
  {Science}\ }\textbf {\bibinfo {volume} {220}},\ \bibinfo {pages} {671}
  (\bibinfo {year} {1983})}\BibitemShut {NoStop}%
\bibitem [{\citenamefont {{\v{C}}ern{\`y}}(1985)}]{Vcerny1985}%
  \BibitemOpen
  \bibfield  {author} {\bibinfo {author} {\bibfnamefont {V.}~\bibnamefont
  {{\v{C}}ern{\`y}}},\ }\href@noop {} {\bibfield  {journal} {\bibinfo
  {journal} {Journal of optimization theory and applications}\ }\textbf
  {\bibinfo {volume} {45}},\ \bibinfo {pages} {41} (\bibinfo {year}
  {1985})}\BibitemShut {NoStop}%
\bibitem [{\citenamefont {Biroli}\ and\ \citenamefont
  {M\'ezard}(2002)}]{Biroli2002}%
  \BibitemOpen
  \bibfield  {author} {\bibinfo {author} {\bibfnamefont {G.}~\bibnamefont
  {Biroli}}\ and\ \bibinfo {author} {\bibfnamefont {M.}~\bibnamefont
  {M\'ezard}},\ }\href {\doibase 10.1103/PhysRevLett.88.025501} {\bibfield
  {journal} {\bibinfo  {journal} {Phys. Rev. Lett.}\ }\textbf {\bibinfo
  {volume} {88}},\ \bibinfo {pages} {025501} (\bibinfo {year}
  {2002})}\BibitemShut {NoStop}%
\bibitem [{\citenamefont {Hunter}\ and\ \citenamefont
  {Weeks}(2012)}]{Hunter12}%
  \BibitemOpen
  \bibfield  {author} {\bibinfo {author} {\bibfnamefont {G.~L.}\ \bibnamefont
  {Hunter}}\ and\ \bibinfo {author} {\bibfnamefont {E.~R.}\ \bibnamefont
  {Weeks}},\ }\href@noop {} {\bibfield  {journal} {\bibinfo  {journal} {Rep.
  Prog. Phys.}\ }\textbf {\bibinfo {volume} {75}},\ \bibinfo {pages} {066501}
  (\bibinfo {year} {2012})}\BibitemShut {NoStop}%
\end{thebibliography}%

\end{document}